\documentclass[hidelinks,11pt]{article}    

\usepackage[left=0.75in,right=0.75in,top=1in,bottom=1in]{geometry}
\usepackage{amsmath, amssymb}
\PassOptionsToPackage{hyphens}{url}\usepackage{hyperref}
\usepackage{graphicx}
\usepackage[singlelinecheck=false, labelsep=space, labelfont=bf]{caption}
\usepackage[labelformat=simple]{subcaption}
\usepackage{tabularx}
\usepackage[round]{natbib}

\DeclareCaptionSubType[roman]{table}
\title{Adjusting for Scorekeeper Bias in NBA Box Scores}
\author{Matthew van Bommel \and Luke Bornn}
\date{Department of Statistics and Actuarial Science, Simon Fraser University}

\begin{document}
	
	\maketitle

\begin{abstract}
Box score statistics in the National Basketball Association are used to measure and evaluate player performance. Some of these statistics are subjective in nature and since box score statistics are recorded by scorekeepers hired by the home team for each game, there exists potential for inconsistency and bias. These inconsistencies can have far reaching consequences, particularly with the rise in popularity of daily fantasy sports. Using box score data, we estimate models able to quantify both the bias and the generosity of each scorekeeper for two of the most subjective statistics: assists and blocks. We then use optical player tracking data for the 2015-2016 season to improve the assist model by including other contextual spatio-temporal variables such as time of possession, player locations, and distance traveled. From this model, we present results measuring the impact of the scorekeeper and of the other contextual variables on the probability of a pass being recorded as an assist. Results for adjusting season assist totals to remove scorekeeper influence are also presented.
\end{abstract}

\section{Introduction}
\label{introduction}
Box score statistics are the baseline measures of performance in the National Basketball Association (NBA). These metrics, either in their raw form or as components of advanced metrics such as player efficiency rating (PER) \citep{hollinger} or win shares (WS) \citep{bbref}, are used by both fans and team officials to help measure and understand player performance. Thus, box score statistics play an influential role in determining playing time, salaries, trade negotiations, marketing potential, and public perception of players, so any inaccuracies or inconsistencies in their attribution can have far reaching impacts. 

Box score statistics for each game are identified and recorded by a team of scorekeepers employed by the home team. Some statistics, such as points scored, are objective and there is little possibility of error by the scorekeepers. However, other statistics, such as assists and blocks, are more subjective in nature. For example, an assist ``is awarded only if, in the judgment of the statistician, the last player's pass contributed directly to a made basket'' \citep{nbau}.  

An example of the consequences of this subjectivity occurred in 1997 when the Vancouver Grizzlies hosted the Los Angeles Lakers. Laker Nick Van Exel was awarded 23 assists, including several that were ``comically bad'', by a disgruntled Grizzlies scorekeeper, in protest of the inaccuracy of box score statistics \citep{craggs09}. The questionable score keeping went undetected, the scorekeeper unpunished, and the recorded box score unaltered. 

While this example is extreme, it demonstrates that inconsistencies in the attribution of box score statistics can occur without notice. Something as simple as scorekeepers having differing views of statistic definitions can affect the comparability of the statistics and thus the perception of player performance. 

With the recent rise in popularity of fantasy sports, these inconsistencies also have monetary implications for participants. FanDuel Inc. and DraftKings Inc., currently the two largest daily fantasy operators in North America, both offer daily fantasy contests for the NBA with point scoring systems relying exclusively on box score statistics \citep{fanduel, draftkings} and participants in the daily fantasy community have noticed the influence of scorekeepers on the scores. In a November 17, 2015 daily fantasy basketball article on ESPN.com, DraftKings analyst Peter Jennings recommends participants select Anthony Davis in the New Orleans Pelicans home game against the Denver Nuggets because ``Davis was much better at home last season (scorekeeper might be a Davis fan) and the trend is continuing" \citep{espn}. 

In this paper we seek to improve the existing methods of examining statistic inconsistency in other sports and apply these methods to NBA data. Our main contributions are the development of specific methods for NBA data, an improved regression model that uses spatio-temporal information provided by optical tracking technology, and a new method of adjusting statistics to correct for inconsistencies. Our adjustment method also allows for the examination of individual scorekeeper accuracy distributions, providing insights into scorekeeper impact on daily fantasy sports. 

The remainder of the paper is organized as follows. In the following section we discuss related work examining statistic inconsistencies in team sports. In Section \ref{assists_and_blocks} we conduct exploratory analysis at the game level into the tendency of scorekeepers to award assists and blocks. Then, in Section \ref{team_adjusted}, we introduce a regression model of assist and block attribution which accounts for the game location, the teams playing, and scorekeeper impacts. This model mirrors the current best practices for estimating the factors influencing the attribution of statistics. Section \ref{new_assists_model} introduces our improvements to the existing methods through a new assist model which takes advantage of optical player tracking data to predict the probability of individual passes being recorded as assists. This section also presents adjusted assist totals which correct for scorekeeper and other biases for a selection of players most affected by the adjustment process. Section \ref{daily_fantasy} examines the impact of scorekeeper inconsistencies on daily fantasy contests. Finally, Section \ref{conclusion} presents conclusions from the results of the paper, as well as a discussion of possible future work and extensions.

\section{Related Work}
\label{related_work}
Regression models are a common tool for sports analytics research, though their application is largely focused within one of two categories: analyzing player performance and predicting win probabilities. Such examinations have spanned several team sports including basketball \citep{deshpande, baghal, fearnhead, basketball_regression, teramoto}, hockey \citep{hockey_regression, macdonald}, baseball \citep{hamrick, neal}, and soccer \citep{groll, oberstone}. 

Regression models have also been employed to examine the effects of biases and inconsistencies in sports. \cite{ref_bias} use NBA box score statistics and game information along with racial data of players, coaches, and referees to examine racial biases of referees. Their models treat referees and the race of players similarly to how our models treat scorekeepers and the home or away status of a team. However, we seek to quantify the behaviour of each scorekeeper individually while \cite{ref_bias} group referees by race. A more applicable model for our objectives was developed by \cite{park_factors} to improve the park factors statistic in Major League Baseball (MLB). They use a regression model to estimate the effects of the design of each park (park factors) on hitting and pitching statistics, while also controlling for a home field advantage and the strength of both teams. These estimated park factor effects are similar to the scorekeeper effects we wish to estimate, except they arise from the unique physical characteristics of each MLB park as opposed to human biases. \cite{schuckers} develop a similar model to estimate the differences in the recording of a number of events across National Hockey League (NHL) rinks. The key difference between this model and the park factors model is that the rink effects model deals with human behaviour and thus includes a rink by home ice interaction effect to capture the bias of the scorekeepers. The estimated rink effects are analogous to our estimated scorekeeper effects and our Model \ref{simple_model} for assists and blocks in the NBA extends this state of the art from hockey and baseball into the domain of basketball. 

In Section \ref{new_assists_model}, we improve upon the existing methods through the inclusion of spatio-temporal covariates, available through optical tracking systems which have recorded data for all NBA courts since the 2013-2014 season. This spatio-temporal information has expanded the scope of possible research, leading to insights that were previously impossible. \cite{cervone} use the spatio-temporal information (including player locations, event annotations, and ball movement) leading up to a given time point to estimate a multiresolution stochastic process model to predict player movements and events after the given time point, with the ultimate goal of computing an expected possession value at any moment in a possession. \cite{basketball_dynamics} use the information in a similar manner to predict the occurrence of near term events, such as a pass or a shot, at a given time point. Aside from the introduction of a novel method of measuring the distance traveled by a player in possession of the ball, our work uses the same features and extraction methods of these previous applications. However, to our knowledge, our work is the first in any sport to use spatio-temporal information to model scorekeeper inconsistencies.

Finally, both \cite{schuckers} and \cite{park_factors} present statistic adjustment methods, which scale the recorded values linearly according to the estimated effects. This is a reasonable adjustment method given their models but since our models contain spatio-temporal covariates, we make use of this additional information and develop a new method to adjust recorded assist numbers over the course of a season. Our adjustment method has the additional advantage of isolating the impact of a variety of effects within the adjustment, providing a more detailed examination of the factors that inflate or deflate statistics.

\section{Assists and Blocks: The Grey Zone of Basketball Analytics}
\label{assists_and_blocks}

According to a former NBA scorekeeper \citep{craggs09}, scorekeepers are given broad discretion over two box score statistics: assists and blocks. Thus, we focus our attention on these metrics. Since assists are highly dependent on the number of made field goals, and blocks dependent on the number of opposing field goal attempts, we examine the assist ratio (AR) and block ratio (BR) rather than the raw totals. Here, the AR and BR for a team are defined as 
\[
\text{AR} = \frac{\text{Team Assists}}{\text{Team Field Goals Made}}
\hspace{20pt}
\text{BR} = \frac{\text{Team Blocks}}{\text{Opponent Field Goals Attempted}}
\]
and can be computed for any given duration, such as a quarter, game, or season.

\begin{figure}
	\centering
	\begin{subfigure}[b]{0.45\textwidth}
		\centering
		\includegraphics[width=\textwidth]{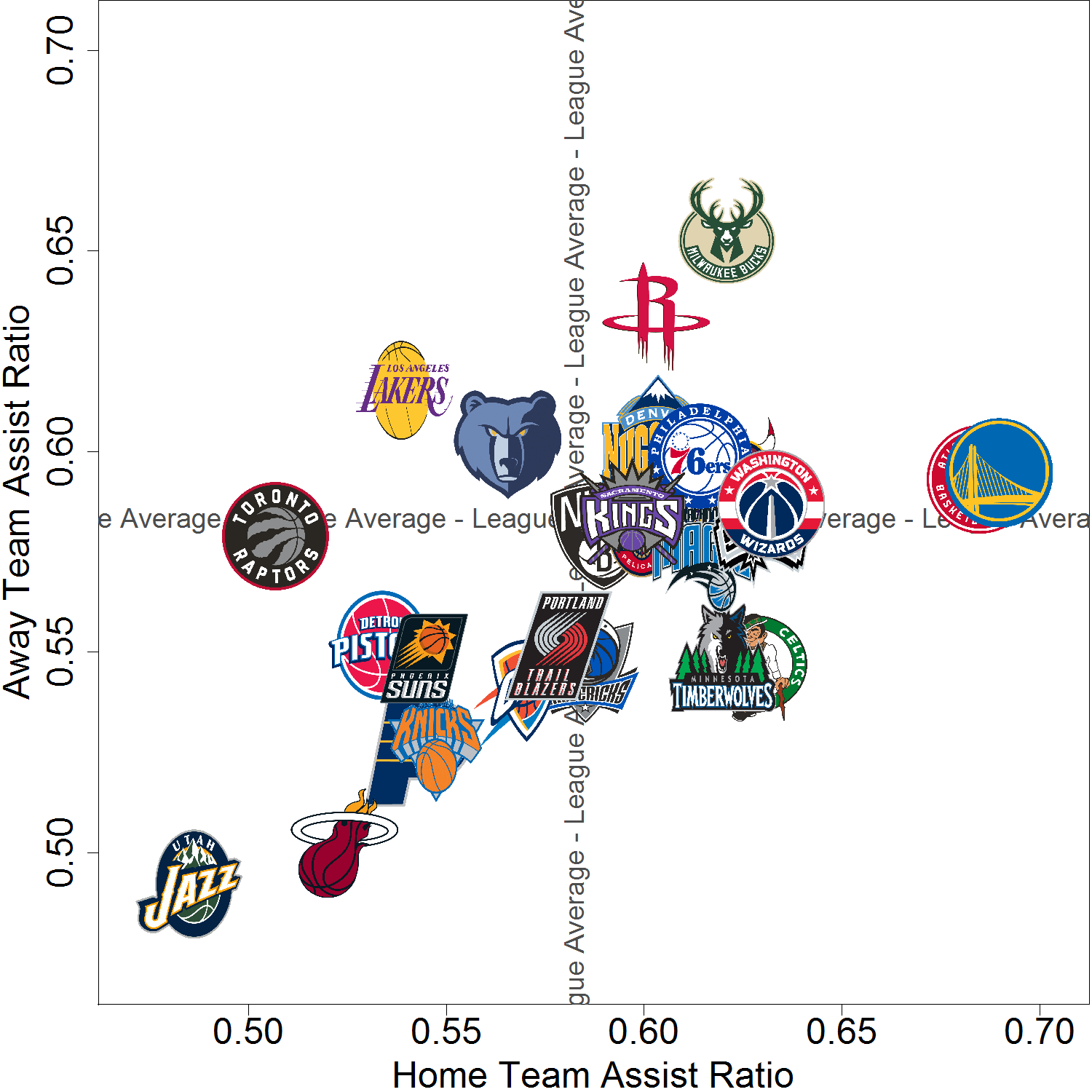} 
	\end{subfigure}
	~
	\begin{subfigure}[b]{0.45\textwidth} 
		\centering
		\includegraphics[width=\textwidth]{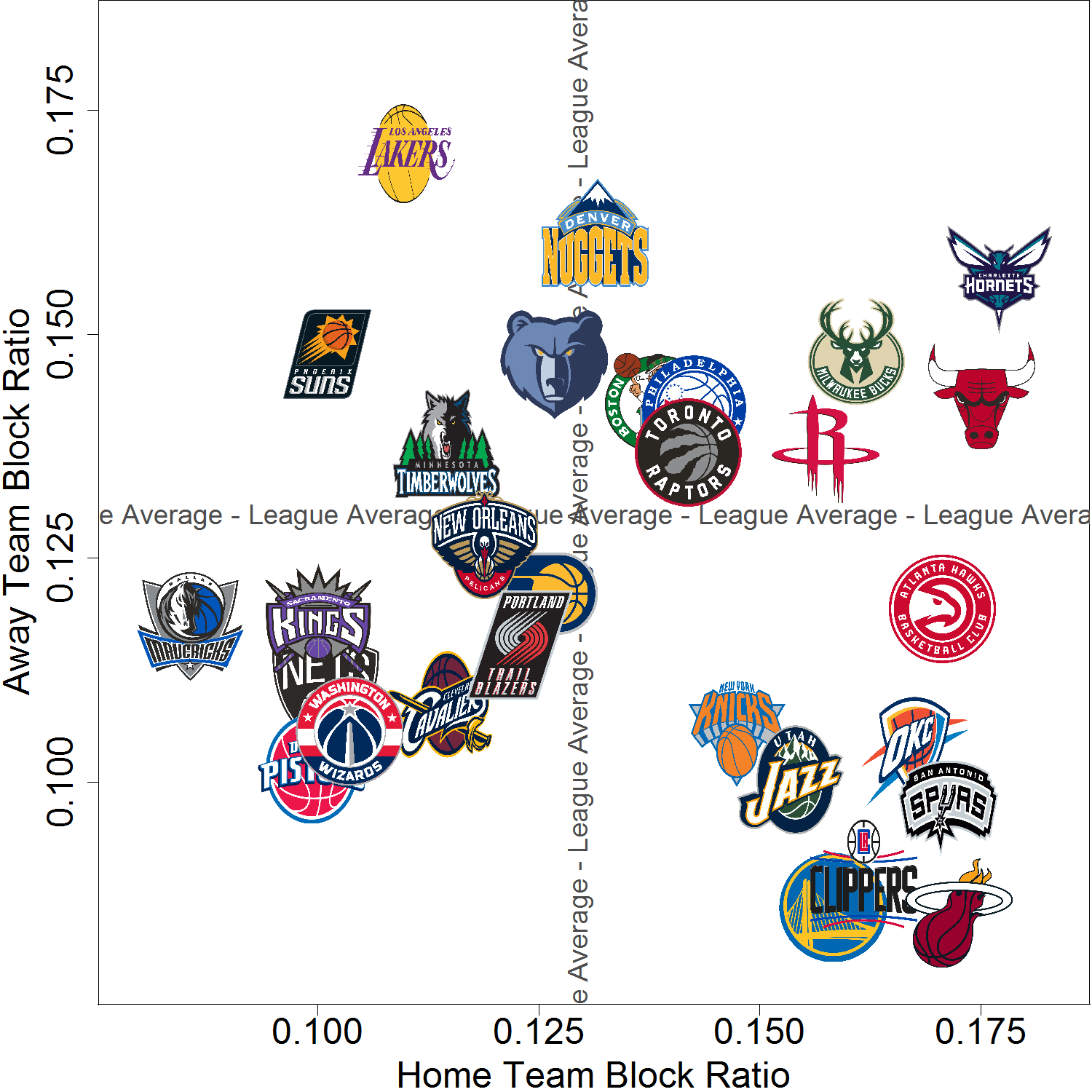} 
	\end{subfigure}	
	\caption{Home team and away team assist ratios (left) and block ratios (right) for the 2015-2016 NBA season for each scorekeeper}
	\label{ar_and_br}
\end{figure}

To examine the scorekeeper impact on these ratios, we use box score data from ESPN.com for the entire 2015-2016 NBA season to compute the season long AR and BR awarded by each scorekeeper to both their home team and the away teams. Figure \ref{ar_and_br} displays the results, demonstrating noticeable differences among scorekeepers. Note that since a scorekeeper is hired by an NBA team to record statistics for all of that team's home games, we use the team names and logos to reference the scorekeepers. Examining assist ratios, many scorekeepers award similar ratios to both home and away teams, however the ratios awarded by some scorekeepers are quite different, either in favour of the home team (Golden State Warriors) or the away team (Toronto Raptors). Similar variability occurs with block ratios with some even more extreme differences favouring either the home team (Miami Heat) or the away team (Los Angeles Lakers). 

Examining the results for the Warriors scorekeeper, it may be the case that they have a bias (either intentional or unintentional) toward their team and thus are more generous when awarding assists. However, it may also be that the Warriors' offensive style focuses on ball movement, resulting in the Warriors making more passes and earning a higher assist ratio than the average NBA team. Similarly, we cannot be certain if the Heat scorekeeper seeks out blocks to award to their team more diligently than the Lakers scorekeeper, or if Hassan Whiteside (a Heat player who led the league in blocks for the 2015-2016 season) is simply that much better of a shot blocker than any player on the Lakers team, such that it inflates the Heat's ratio to well above that of the Lakers. In the following section, we reduce this uncertainty through models of AR and BR which separate the influence of the teams from the influence of the scorekeepers.

\section{Team Adjusted Models}
\label{team_adjusted}

We now introduce a regression model which examines the factors influencing the AR or BR for a single team in a given game. Aside from the actions of the scorekeeper, we suspect the primary variables influencing the assist or block ratio awarded by a scorekeeper in a game are the teams in that game. Both the style of play and the skill of the players on a team influence its likelihood to record an assist or block. Similarly, the style and player skill of a team also influence its likelihood of allowing an assist or block. Thus, the models estimate a ``team'' ($\pmb{\beta}_T$) and ``opponent'' ($\pmb{\beta}_O$) effect for each team, corresponding to the likelihood of each team to respectively record or allow a given statistic. Our models also include a ``home'' ($\beta_H$) effect, common to all teams. This effect is present only when the team of interest is the home team and represents any possible league-wide home court advantage resulting in increased performance with respect to the given statistic. 

The final two non-intercept effects estimated in our models are the ``scorekeeper generosity'' ($\pmb{\beta}_G$) and ``scorekeeper bias'' ($\pmb{\beta}_B$) effects corresponding to each team's scorekeeper. $\pmb{\beta}_G$ measures how likely a scorekeeper is to award assists or blocks to both teams while $\pmb{\beta}_B$ measures how much more likely a scorekeeper is to award an assist or block to the home team compared to the away team. Isolated from the influence of the other previously mentioned effects, these effects will provide insight into the consistency of scorekeepers across the NBA.

Let $\text{R}_{i}$ be the expected ratio of interest (either AR or BR) for a given team-game combination $i$. Our model for $\text{R}_{i}$ takes the form
\begin{equation}
\label{simple_model}
\text{R}_{i} = \beta_0 + H_{i} \beta_H + \text{T}_{i} \pmb{\beta}_T + \text{O}_{i} \pmb{\beta}_O + \text{S}_{i} \pmb{\beta}_{G} + \text{S}'_{i} \pmb{\beta}_{B}
\end{equation}
where $\text{H}_{i}$ is an indicator variable denoting if the team in $i$ is the home team, $\text{T}_{i}, \text{O}_{i}$, and  $\text{S}_{i}$ are each $30 \times 1$ indicator (one-hot encoded) vectors for the team, its opponent, and the scorekeeper for team-game combination $i$ respectively, and $\text{S}'_{i} = \text{S}_{i} \times \text{H}_{i}$ is a $30 \times 1$ indicator vector denoting the scorekeeper if the team in team-game combination $i$ is the home team (and is a zero vector otherwise). These variable definitions are summarized in Table \ref{initial_parameters}. Additionally, $\beta_0$ and $\beta_H$ are estimated coefficients and $\pmb{\beta}_T, \pmb{\beta}_{G},$ and $\pmb{\beta}_{B}$ are $1 \times 30$ row vectors of estimated coefficients. The coefficients measuring scorekeeper effects ($\pmb{\beta}_{G}$ and $\pmb{\beta}_{B}$) are the coefficients of interest while the remaining coefficients account for the impact of other influential factors.  

Model \ref{simple_model} is essentially the model presented by \cite{schuckers} applied to NBA statistics, but with two main differences. First, we introduce two scorekeeper effects to mirror their single rink effect. We believe there are two distinct potential differences in scorekeeper behaviour and thus dividing the scorekeeper effect provides additional insight. Second, we model ratios of statistics rather than simple counts. Using ratios focuses the model on the scorekeepers' decisions rather than the teams' ability to generate shots, since assists and blocks are dependent on made and missed field goals respectively. Since we do not examine count data, we also do not employ the logarithmic transformation used by \cite{schuckers}.

\begin{table}
	\caption{Variables for the team adjusted models}
	\label{initial_parameters}
	\begin{tabularx}{\textwidth}{ l X }
		\hline\noalign{\smallskip}
		\textbf{Notation} & \textbf{Definition} \\
		\noalign{\smallskip}\hline\noalign{\smallskip}
		$H_i$ & $2 \times 1$ indicator vector denoting whether the team is home or away \\
		$T_i$ & $30 \times 1$ indicator vector denoting the team \\
		$O_i$ & $30 \times 1$ indicator vector denoting the opponent \\
		$S_i$ & $30 \times 1$ indicator vector denoting the scorekeeper \\
		$S'_i$ & $30 \times 1$ indicator vector denoting the interaction of home and scorekeeper ($H_i$ and $S_i$) \\
		\hline\noalign{\smallskip}
	\end{tabularx}
\end{table}

\begin{figure*}
	\centering
	\begin{subfigure}[b]{0.45\textwidth}
		\centering
		\includegraphics[width=\textwidth]{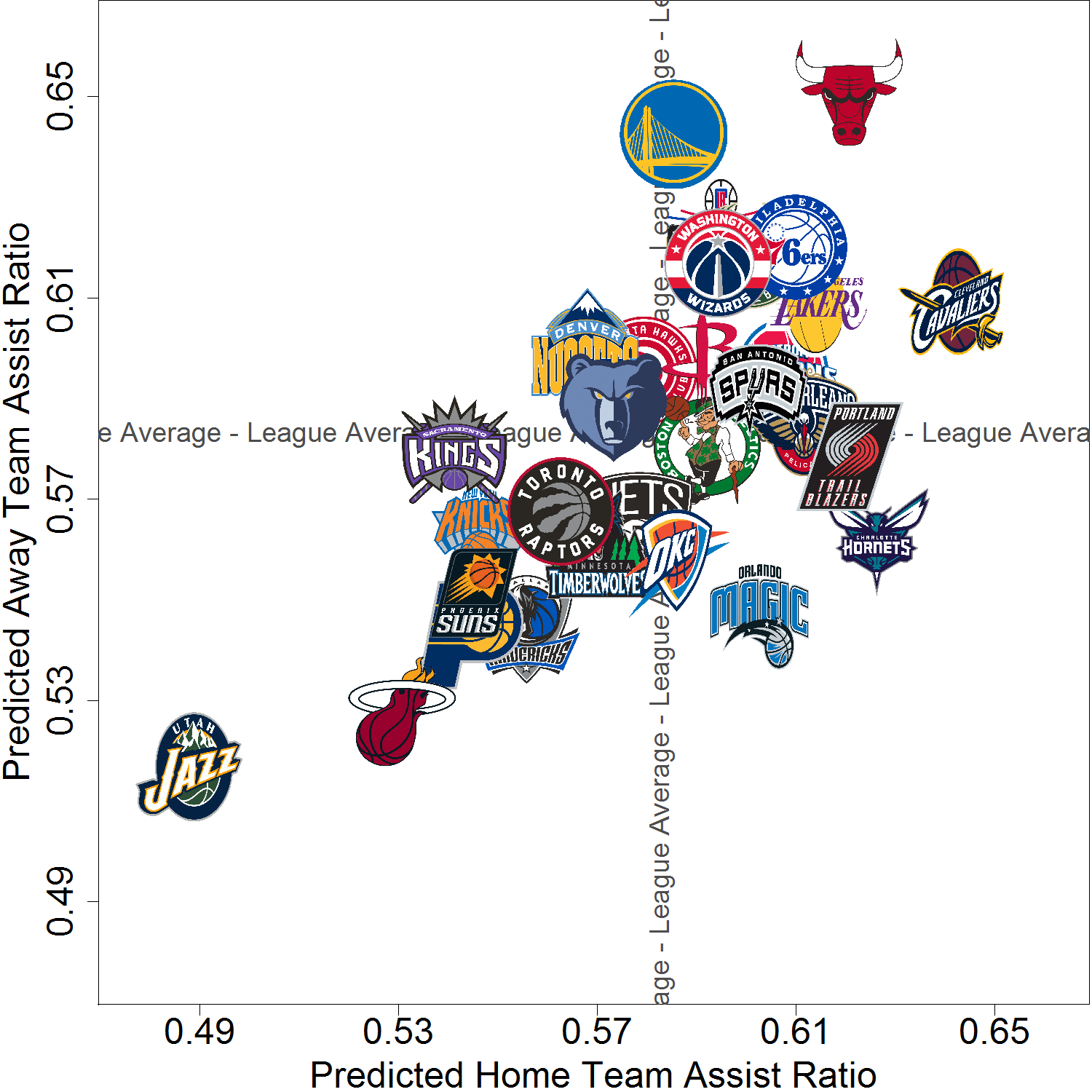} 
	\end{subfigure}
	\begin{subfigure}[b]{0.45\textwidth} 
		\centering
		\includegraphics[width=\textwidth]{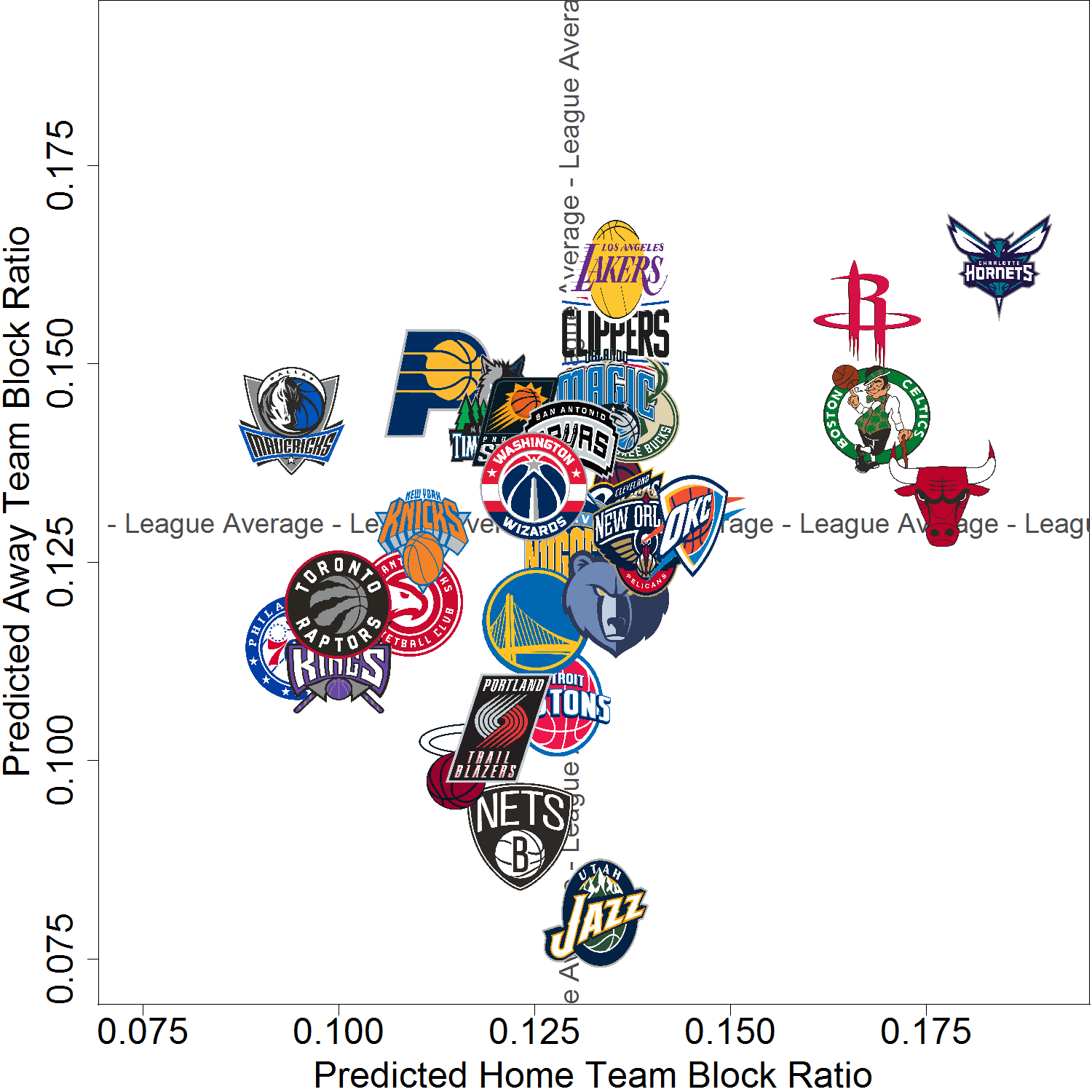} 
	\end{subfigure}	
	\caption{Predicted assist ratios (left) or block ratios (right) awarded by each scorekeeper to the home team and an unspecified away team based on the coefficients of the team-level model}
	\label{sk_impact_on_ar_br}
\end{figure*}

To estimate the model parameters, we again use box score data from ESPN.com for the entire 2015-2016 NBA season, but this time we compute the assist and block ratios of each team in every game. To compare the consistency of the 30 scorekeepers for each ratio, we compute predicted ratios awarded by each scorekeeper to both the home team and an unspecified away team. Let the predicted home and away team ratios (either AR or BR) for scorekeeper $s$ be denoted $\text{PR}_{H_s}$ and $\text{PR}_{A_s}$ respectively. Then
\begin{align*}
\text{PR}_{H_s} &= \overline{\text{LR}} + \left( \pmb{\beta}_{G}^{(s)} - \bar{\pmb{\beta}}_{G} \right) + \left( \pmb{\beta}_{B}^{(s)} - \bar{\pmb{\beta}}_{B} \right) \\
\text{PR}_{A_s} &= \overline{\text{LR}} + \left( \pmb{\beta}_{G}^{(s)} - \bar{\pmb{\beta}}_{G} \right)
\end{align*}
where $\overline{\text{LR}}$ is the season long league ratio (AR or BR), the $\pmb{\beta}^{(s)}$ are the entries in the $\pmb{\beta}$ vectors corresponding to scorekeeper $s$, and the $\bar{\pmb{\beta}}$ are the average of the elements in the $\pmb{\beta}$ vectors. The resulting $\text{PR}_{H_s}$ and $\text{PR}_{A_s}$ values for all 30 scorekeepers for both AR and BR are presented in Figure \ref{sk_impact_on_ar_br}. Note that the scaled $\pmb{\beta}_{G}$ values are the differences between the league average and the away team predicted ratios while the scaled $\pmb{\beta}_{B}$ values can be determined by subtracting the away team predicted ratios from the home team predicted ratios. It can be noticed that some of the observations from Figure \ref{ar_and_br} still hold. For example, both AR figures indicate the Utah Jazz scorekeeper is unbiased but not generous. However, there are also substantial differences between the figures. The BR results for the Atlanta Hawks and Sacramento Kings scorekeepers are nearly on opposite ends of the home team block ratio range in Figure \ref{ar_and_br}, but in Figure \ref{sk_impact_on_ar_br}, the two scorekeepers have very similar results.

The model results appear to confirm several interesting characteristics. First, some scorekeepers are biased against their home team. Away teams are much more likely to be awarded an assist by the Jazz scorekeeper or a block from the Dallas Mavericks scorekeeper than the corresponding home teams are. Also, the effect of the scorekeeper on different statistics is not necessarily consistent. The Philadelphia 76ers scorekeeper is among the most likely to award an assist to either team but is among the least likely to award a block, particularly to the home team. 

However, when examining how well the models fit the data we see that the AR model and BR model have coefficients of determination ($R^2$ values) of 0.279 and 0.228 respectively. Thus, while the results provide some indication of the factors influencing the attribution of statistics, there is certainly room for improvement. Additionally, since the scorekeeper bias effect for a team's scorekeeper is only present when a team is home and there is only a single home effect common to all teams, it is not clear if the scorekeeper bias effect truly measures scorekeeper bias or if it measures a team specific home effect. It may be the case that the average pass made by a team at home is of a different quality than passes made by that team away from home. In the following section, we focus on assists and introduce a new model which uses spatio-temporal information to improve the results and reduce the potential of confounding effects.

\section{A New Assist Model: Adjusting for Context}
\label{new_assists_model}
While assists and blocks are both subjective statistics, the factors involved in their attribution differ. Blocks occur in an instant (the moment that a defender makes contact with an opposing player's shot) while assists involve two components (a pass and a made basket) which can develop over the course of several seconds and can include additional actions such as pivots, dribbles, and defender movements. Thus the context surrounding an assist is fundamental to its attribution. This section takes advantage of this context to build a contextual assist model that improves upon the currently available methods.

For this new model, we narrow our focus to the level of individual passes and examine passes with the potential to be recorded as assists. We define a potential assist to be a completed pass from a passer to a shooter who then scores a field goal within seven seconds of receiving the pass. The shooter is permitted to dribble and move after receiving the pass, as long as he maintains possession of the ball until the successful shot (no rebounds, turnovers, or additional passes may occur). Note that while an inbounds pass can be credited as an assist, for simplicity we will only examine passes made while the ball was in play.   

\subsection{Extracting Spatio-Temporal Features}
\label{spatio-temporal_features}
\begin{figure*}
	\centering
	\includegraphics[width=0.6\textwidth]{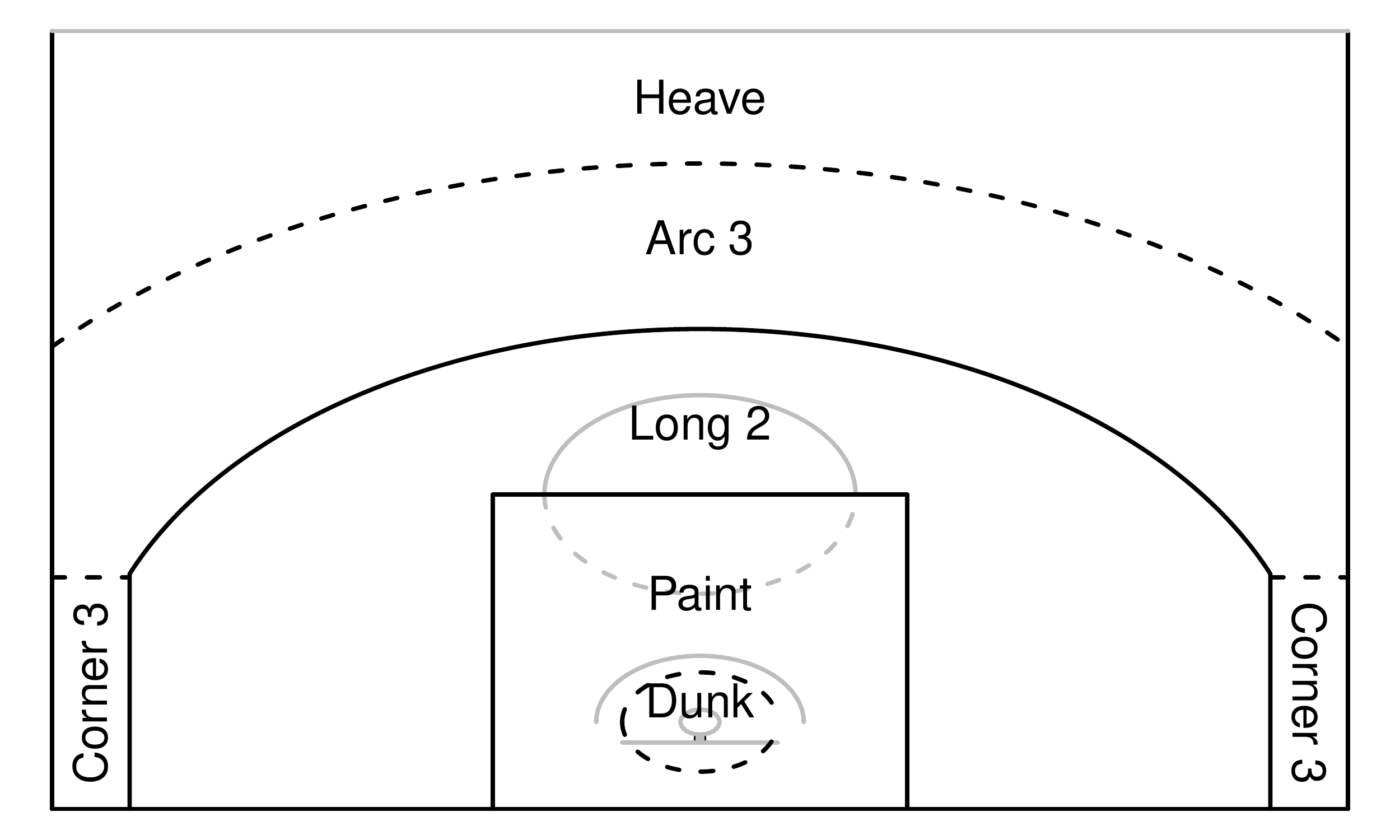}
	\caption{The offensive half court divided into six court zones: Dunk, Paint, Long 2, Arc 3, Corner 3, and Heave. The gray lines are court lines that do not divide the zones, the solid black lines are court lines that divide the zones, and the dashed black lines divide the zones but are not found on the court. The gray line surrounding the Dunk zone forms a circle of radius three feet centered at the center of the basket, the  gray lines separating the Corner 3 and Arc 3 zones are drawn horizontally from the sidelines to the points at which the three point line begins to curve, and the dividing line between the Arc 3 and Heave zones is drawn ten feet beyond the three point arc. Note that the Heave zone extends beyond the half court line and covers the remainder of the court.}
	\label{court_zone_figure}
\end{figure*}

When examining an individual potential assist, there are many contextual spatio-temporal factors that influence its probability of being recorded as an assist by the scorekeeper. Characteristics of the shooter's possession, such as possession length, number of dribbles, and distance traveled, are particularly relevant to the determination of assists, since the pass must be considered to lead directly to the made field goal. The locations of the passer and shooter may also influence the probability of a recorded assist. In order to measure these location impacts, we divide the court into the 6 distinct zones displayed in Figure \ref{court_zone_figure}.

To measure the contextual variables for potential assists, we use SportVu optical player tracking data from STATS LLC which contains the X- and Y-coordinates of each of the 10 players on the floor and X-, Y-, and Z-coordinates of the ball, which are recorded 25 times per second throughout the course of a game. Annotations for events including passes, dribbles, and shots are also included, as well as additional information for player and team identification, dates and times, and game clock and shot clock times. The data set contains game data for 1227 of the 1230 2015-2016 NBA regular season games and all teams have at least 81 of their 82 total games included. We examine the 82,493 potential assists contained in the data set, of which, 54,111 (65.59\%) were recorded assists. 

\begin{figure*}
	\centering
	\begin{subfigure}{0.49\textwidth}
		\centering
		\caption{}
		\includegraphics[width=\textwidth]{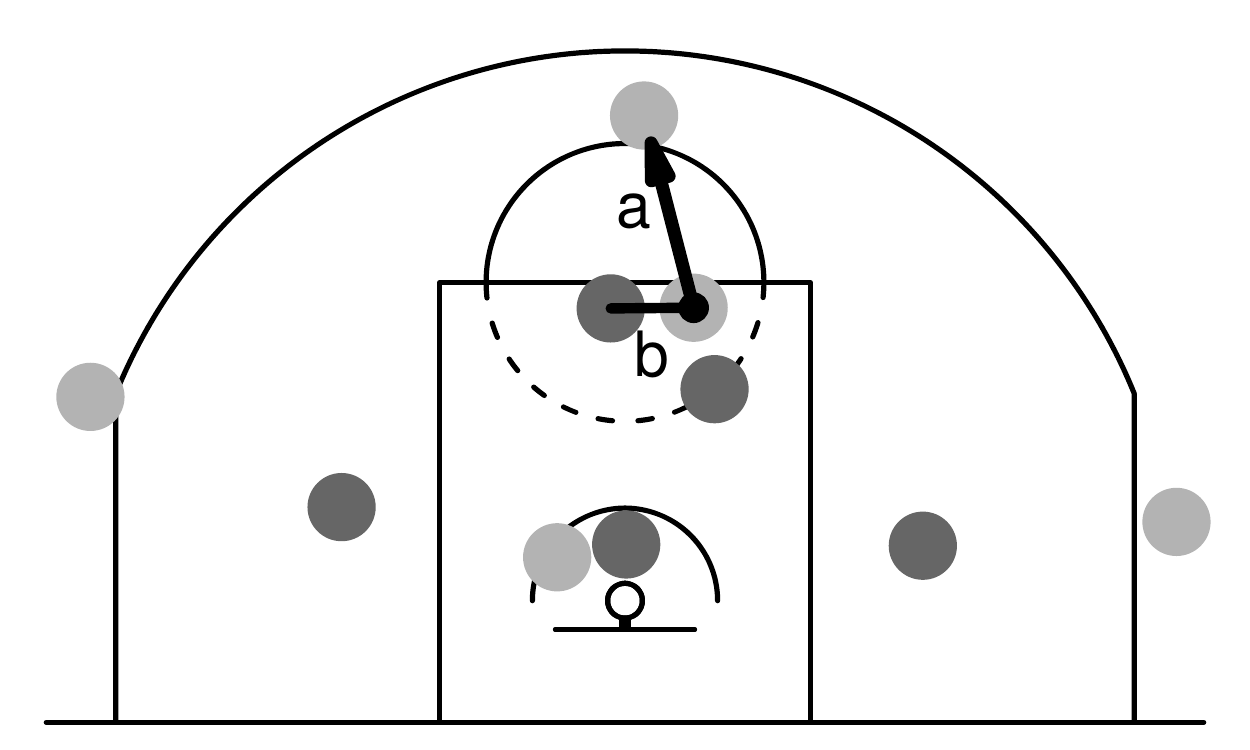} 
	\end{subfigure}	
	\begin{subfigure}{0.49\textwidth} 
		\centering
		\caption{}
		\includegraphics[width=\textwidth]{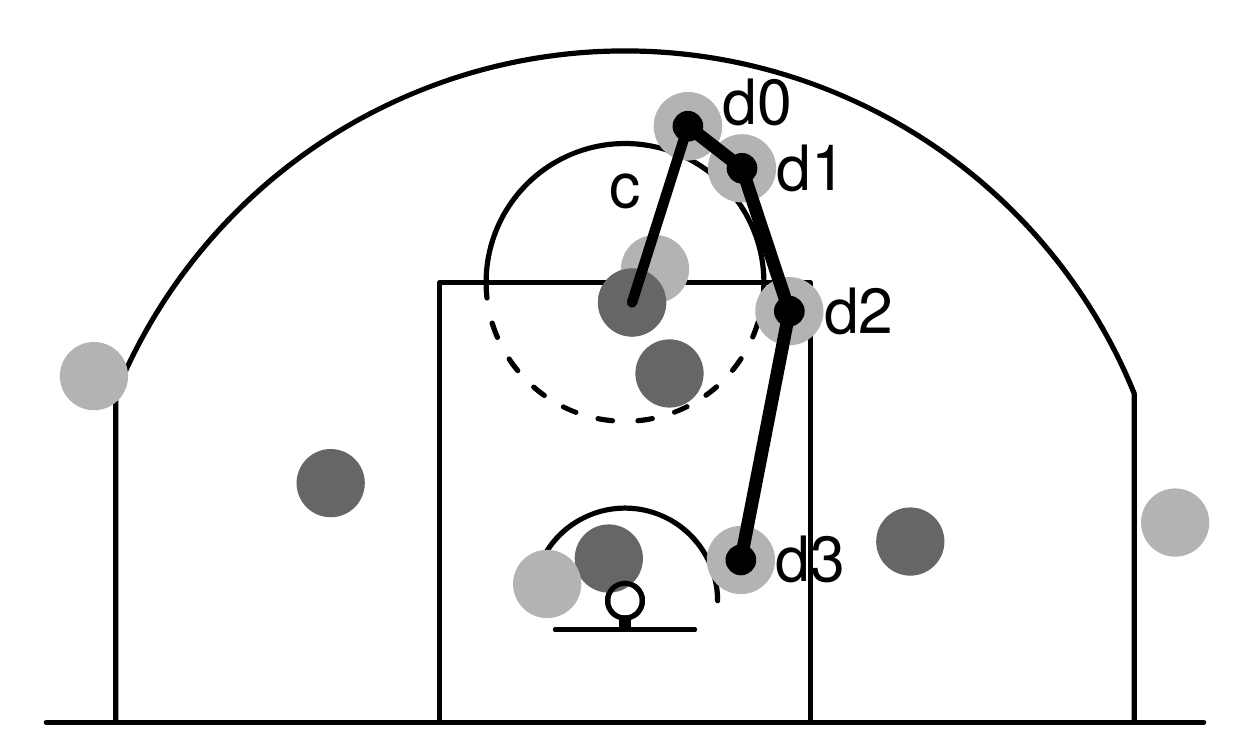}
	\end{subfigure}	
	\caption{Optical tracking data for the stages of a potential assist which occurred in the first quarter of a January 7, 2015 game featuring the Los Angeles Clippers hosting the Los Angeles Lakers. Each point represents one of the 10 players on the floor, with the lighter points representing Clippers players (on offense) and the darker points representing Lakers players (on defense). The ball is marked by the black dot on top of the player who has possession. In \textbf{i}, Clippers guard Chris Paul is in possession of the ball in the $C_{(7)}=$ Paint court zone and is about to make a pass of distance $C_{(4)} = 11.09$ ft along arrow \textbf{a} to Clippers forward Blake Griffin. At the moment of the pass, the distance of the nearest defender to Paul is $C_{(5)} = 3.58$ ft, measured by line \textbf{b}. In \textbf{ii}, Griffin receives the pass in the $C_{(8)}=$ Long 2 court zone. At this moment, the distance of the nearest defender to Griffin  is $C_{(6)} = 13.63$ ft, measured by line \textbf{c}. After receiving the pass (\textbf{d0}), Griffin drives to the basket, takes $C_{(2)} = 2$ dribbles (\textbf{d1} and \textbf{d2}), and shoots the ball (\textbf{d3}) (the locations of the other players during the drive are held constant for simplicity). During this drive, Griffin travels $C_{(3)} = 20.41$ ft over $C_{(1)} = 1.82$ seconds. On this play, Griffin's shot attempt was successful and the Clippers scorekeeper awarded Paul an assist.}
	\label{sample_play}
\end{figure*}

The spatio-temporal context variables that will be included in our new contextual model are presented in Table \ref{context_parameters}. Note that the set of event annotations in the data set mark when a player releases or receives a pass and when the ball is dribbled or released for a shot. Thus, the methods of obtaining values for $C_{(1)}, C_{(2)}, C_{(4)}, C_{(7)}, C_{(8)},$ and $C_{(9)}$ are straightforward. Since the SportVu location data is noisy, summing the distance between each observation for a player's location over the range of time that player is in possession of the ball is likely to overestimate the total distance traveled by that player. To correct for this, we take advantage of the NBA traveling violation which prevents a player in control of the ball to move without dribbling the ball (with the exception of pivoting or of two steps allowed immediately after receiving a pass or concluding a final dribble) and define $C_{(3)}$ to be sum of the distances between the observations of a player's location when he receives possession of the ball, each time he dribbles the ball, and when he releases a shot. Finally, $C_{(5)}$ and $C_{(6)}$ are determined by calculating the distance between the corresponding offensive player and each of the five opposing players on the court (at the defined moment in time) and taking the minimum of those five distance values. The stages of a sample potential assist are displayed in Figure \ref{sample_play} to demonstrate the computation of the parameters in Table \ref{context_parameters}.  

\begin{table}
	\caption{Spatio-temporal variables for the contextual model}
	\label{context_parameters}
	\begin{tabularx}{\textwidth}{ l X }
		\hline\noalign{\smallskip}
		\textbf{Notation} & \textbf{Definition} \\
		\noalign{\smallskip}\hline\noalign{\smallskip}
		$C_{(1)}$ & Continuous variable denoting the time (seconds) of the shooter's possession \\
		$C_{(2)}$ & Discrete variable denoting the number of dribbles taken during the shooter's possession \\
		$C_{(3)}$ & Continuous variable denoting the distance (feet) traveled by the shooter during possession \\
		$C_{(4)}$ & Continuous variable denoting the distance (feet) traveled by the pass \\
		$C_{(5)}$ & Continuous variable denoting the distance (feet) of the nearest defender to the passer at the time of the pass \\
		$C_{(6)}$ & Continuous variable denoting the distance (feet) of the nearest defender to the shooter at the start of the shooter's possession \\
		$C_{(7)}$ & $6 \times 1$ indicator vector denoting the court zone corresponding to the passer at the time of the pass \\
		$C_{(8)}$ & $6 \times 1$ indicator vector denoting the court zone corresponding to the shooter at the start of the shooter's possession \\
		$C_{(9)}$ & $36 \times 1$ indicator vector denoting the interaction of passer and shooter court zones ($C_{(7)}$ and $C_{(8)}$) \\
		\hline\noalign{\smallskip}
	\end{tabularx}
\end{table}

\subsection{Estimating the Contextual Assist Model}
\label{estimating_new_assist_model}
For the contextual assist model, we use logistic regression to predict the probability of the $j^{th}$ potential assist being recorded as an assist, given a variety of contextual factors. The contextual model takes the form
\begin{equation}
\label{contextual_model}
P(A_j = 1) = \sigma 
\begin{pmatrix}
\beta^*_0 + \text{H}_{j} \beta^*_H + \text{T}_{j} \pmb{\beta}^*_T + \text{O}_{j} \pmb{\beta}^*_O + \text{S}_{j} \pmb{\beta}^*_{G} + \text{S}'_{j} \pmb{\beta}^*_{B} \\
+ \text{N}_j \pmb{\beta}^*_N + \text{P}_j \pmb{\beta}^*_P + \displaystyle\sum_{k=1}^{9} C_{(k)j} \pmb{\beta}^*_{C_k} 
\end{pmatrix}
\end{equation}
where $\sigma(x) = \exp(x)/(1+\exp(x))$ and $A_j$ is an indicator function equal to 1 when potential assist $j$ is a recorded assist. The terms common to the team-level model (Model \ref{simple_model}) share the same definitions as outlined in Table \ref{initial_parameters} except the index $j$ refers to a single potential assist achieved in a given team-game combination. For the new model terms, $N_j$ is an indicator vector denoting the name of the passer (from the 486 unique passers in the data), $P_j$ is an indicator vector denoting the primary position (point guard, shooting guard, small forward, power forward, or center) of the passer, and the $C_{(k)j}$ variables are the additional spatio-temporal context variables defined in Table \ref{context_parameters}. Finally, $\beta^*_0$ and $\beta^*_H$ are estimated coefficients and the $\pmb{\beta}^*_l$ coefficient vectors are $1 \times n_l$ row vectors of estimated coefficients, where $n_l$ is the number of rows of the observation vector which is multiplied by the corresponding coefficient vector. In estimating the model, we use logistic regression with an L2 penalty on the $\beta$ coefficients, learned through cross-validation. Thus, the model estimation solves
\[
\min_{\pmb{\beta}} \left[ - \left( \frac{1}{N} \sum^N_{j=1} A_j M - \log(1 + e^{M}) \right) + \lambda ||\pmb{\beta}^*||^2 \right]
\]
over a grid of values of $\lambda$ covering the range of interest where $\pmb{\beta}^*$ is a vector of all $\beta^*$ coefficients estimated in the contextual model, $N=82,493$ (the number of potential assists in the data set), and $\sigma(M)$ is the right hand side of Equation \ref{contextual_model} describing the contextual model. 

The inclusion of contextual information in the model fixes a key issue of Model \ref{simple_model}. Using Model \ref{simple_model}, it is impossible to isolate the impacts of scorekeeper bias and specific home team effects within the scorekeeper bias effect, since passes made by some or all home teams may differ in quality from passes made by those teams when they are away from home. However, with the inclusion of measures of pass quality within the model (nearest defender distance to shooter when the pass is received, number of dribbles and distance traveled by shooter to attempt a shot, pass location, and pass distance), the only possible confounding effects are other measures of pass quality not included in the model. Therefore, we can confidently assume that true scorekeeper bias is the main factor influencing the estimated scorekeeper bias effect.

Our choice of an L2 penalty differs from that of \cite{hockey_regression} who choose an L1 penalty for their logistic regression model estimating player contribution in hockey. Their decision is based on the variable selection benefits of the L1 penalty. However, the covariates of their model are limited to teams and players while our model contains additional covariates, including several contextual covariates that are highly correlated (such as number of dribbles and distance traveled). Because of this, an L2 penalty is a natural choice for its improved predictive performance in the presence of collinearity \citep{tibshirani}.

It is important to note that the results of the contextual assist model depend on the selected value of $\lambda$. In particular, the estimated coefficients for variables with relatively few observations (such as some passer effects in $\pmb{\beta}^*_N$ and some court zone interaction effects in $\pmb{\beta}^*_{C_9}$) are more sensitive to shrinkage. Thus, while the relative order of the resulting effects associated with these coefficient values are largely unaffected, the magnitudes of the effects are impacted by the selected $\lambda$. In order to mitigate this impact we use 100-fold cross validation to select the optimal $\lambda$ value and use this value to estimate the model used to generate the results presented in the remainder of this paper.

\subsection{Contextual Assist Model Results}
\label{new_assist_model_results}
The primary focus of this section is to examine the results corresponding to the new variables introduced to the contextual model that were not included in the team-level model. However, we first examine the impact of these additional variables on the scorekeeper effects. Comparing the $\pmb{\beta}_{G}$ and $\pmb{\beta}_{B}$ coefficients of the team-level model to the $\pmb{\beta}^*_{G}$ and $\pmb{\beta}^*_{B}$  coefficients of the contextual model, both pairs of estimated coefficients are positively correlated. However, the correlation of $\pmb{\beta}_{G}$ and $\pmb{\beta}^*_{G}$ is 0.892 compared to only 0.597 for $\pmb{\beta}_{B}$ and $\pmb{\beta}^*_{B}$. One possible explanation for this difference is that the range of generosity coefficient values compared to the bias values is much greater in the contextual model. Thus, it may be the case that the variation explained by $\pmb{\beta}^*_{B}$ is better explained by $\pmb{\beta}^*_{G}$ or other new coefficients in the contextual model. Another possible explanation is that the bias values are estimated using fewer observations (since bias coefficients apply only to the home team of each game and the generosity coefficients apply to both teams) making them less reliable compared to the generosity values.

Shifting focus to the new variables introduced in the contextual model, we can isolate the impact of a single variable on the probability of a potential assist being recorded as an assist by examining an average potential assist, and observing how its recorded assist probability changes as we manipulate the value of the variable of interest. An average potential assist is a potential assist with no impact from the indicator variables or vectors, and the average values over all potential assists of the other variables. Thus an average potential assist is a pass that travels 18.21 feet from a passer whose nearest defender is 6.67 feet away, to a shooter whose nearest defender is 9.63 feet away when he catches the ball and who then takes 1.87 dribbles, traveling 16.00 feet, over 2.59 seconds before scoring a field goal. The model predicts that this average potential assist has a 39.23\% chance of being a recorded assist. 

Let $V$ be the sum of the estimated intercept coefficient and the average potential assist values multiplied by their corresponding estimated coefficient values, that is $V = \beta^*_0 +  2.59 \beta^*_{C_1} + 1.87 \beta^*_{C_2} + 16.00 \beta^*_{C_3} + 18.21 \beta^*_{C_4} + 6.67 \beta^*_{C_5} + 9.63 \beta^*_{C_6}$. Also, let $I$ be any variable of interest from the contextual assist model, with corresponding estimated coefficient vector $\pmb{\beta}_I^*$. If $I$ is a variable included in $V$, redefine $V$ without the corresponding variable term. The influence of a given value $I^*$ of the variable of interest $I$ has the following effect ($E$) on the probability of an average potential assist being recorded as an assist:
\begin{equation}
E = \sigma \left(V + I^* \pmb{\beta}_I^* \right) - \sigma \left(V\right)
\label{effect_computation}
\end{equation}
where $E$ is measured in units of change in probability. 

\begin{figure*}
	\centering
	\includegraphics[width=\textwidth]{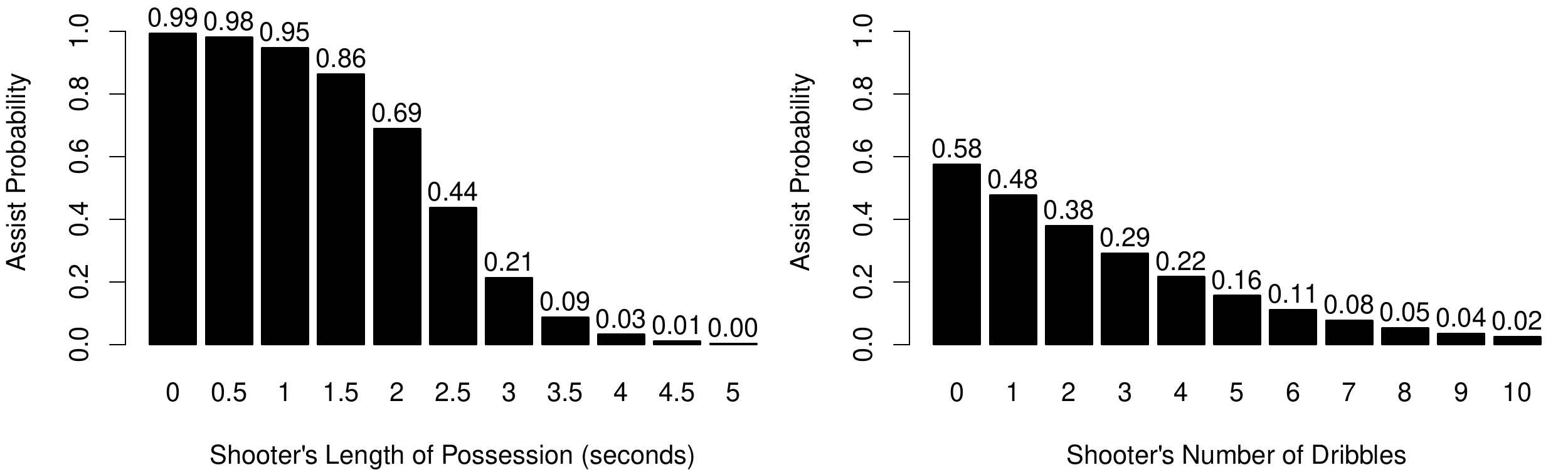}
	\caption{Select predicted recorded assist probabilities of the average potential assist as a function of the continuous possession length (left) or the discrete number of dribbles (right), computed using Equation \ref{effect_computation}}
	\label{time_dribbles}
\end{figure*}

We first use the above effect computation expression to examine the impact of some of the contextual variables in the model. Figure \ref{time_dribbles} demonstrates the effects of changing the possession length or the number of dribbles of the average potential assist. While both variables effect the probability of a potential assist being a recorded assist, possession length has a more substantial effect, as demonstrated by the wider range of probabilities and the more drastic decrease in probability as the value of the variable increases.

\begin{figure*}
	\centering
	\includegraphics[width=\textwidth]{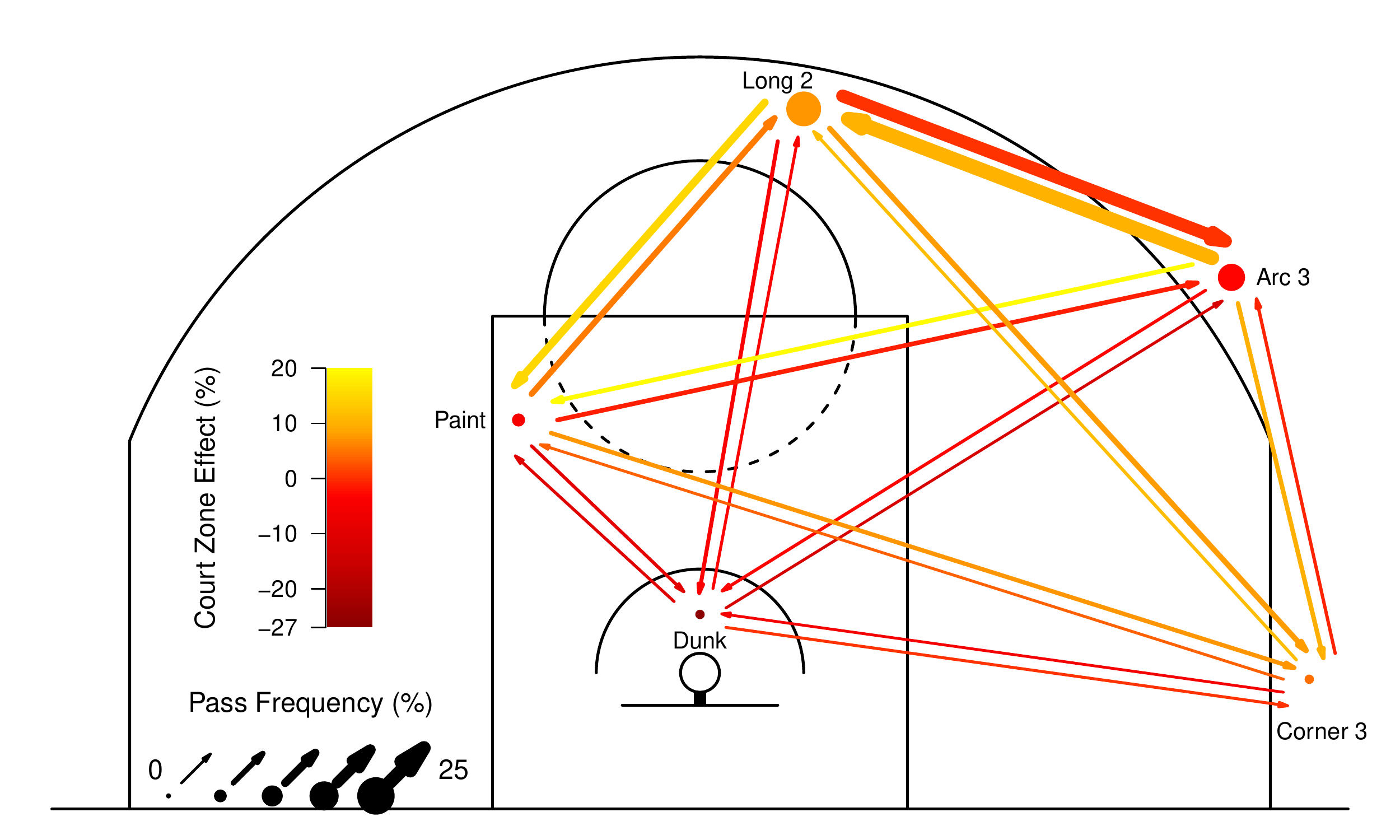}
	\caption{The impact of passer and shooter location on the recorded assist probability of an average potential assist and the frequency of passes between each location pair. For each pair of zones, the arrow points in the direction of the pass. Note that the point in each zone represents passes made within that zone.}
	\label{court_zone_results}
\end{figure*}

We also examine the impact of the passer and shooter locations. Here we ignore passes to or from the Heave zone as they have relatively low probabilities of being recorded assists. The resulting impact of each remaining pair of zones, and the frequency of passes between them, are presented in Figure \ref{court_zone_results}. Accounting for the other contextual model factors (including player positions, which are discussed in the following paragraph), for passer locations in zones closer to the basket (Paint and Dunk), passes to the Corner 3 zone are the most likely to be recorded assists. For the Long 2 and Arc 3 zones, passes to the Paint zone are those most likely to be recorded assists. For four of the zones, passes to the Dunk zone are the least likely to lead to assists (for the Arc 3 zone, passes within that zone are slightly less likely to be recorded assists). Overall, passes from the Arc 3 zone to the Paint zone and passes within the Dunk zone are respectively the most and least likely to be recorded as assists.

Next, we shift our focus to the non-contextual variables of the contextual model, beginning with the primary position of the passer. The results of the position effects are displayed in Table \ref{position_effects} and show that even after accounting for all contextual variables in the model, the probability of an average potential assist being a recorded assist is greatest for point guards and least for centers, with a 7.77\% difference in probabilities between the two positions. We propose two possible explanations for this discrepancy. First, the average pass by a point guard may have a higher probability of being a recorded assist due to characteristics beyond the scope of the contextual model. However, if the model captures all important characteristics, then the discrepancy may be the result of bias from scorekeepers with respect to passer positions. 

\begin{table}
	\caption{Passer position effects where ``Effect'' is the isolated effect of the passer position on the recorded assist probability of an average assist and is computed using Equation \ref{effect_computation}}
	\label{position_effects}
	\begin{tabular}{l c c c c c}
		\hline\noalign{\smallskip}
		\textbf{Position} & Point Guard & Shooting Guard & Small Forward & Power Forward & Center \\
		\noalign{\smallskip}\hline\noalign{\smallskip}
		\textbf{Effect (\%)} & 3.76 & 0.59 & -0.28 & -0.33 & -4.01 \\
		\hline\noalign{\smallskip}
	\end{tabular}
\end{table}

Continuing with the non-contextual variables, we end the results examination of the contextual model with the impact of the passer on the probability of a recorded assist. The results for the top and bottom ten passer effects are displayed in Table \ref{passer_effects}. Since passer positions are accounted for separately, they should not be the primary influence of passer coefficients. This idea holds true in the results as both the top and bottom ten include players from all five positions. The difference in effects between the player with the highest effect (Collison) and the player with the lowest effect (Roberson) is a substantial 27.41\%. Similarly to the position effect, we suspect the differing passer effects are the result of either characteristics of passes not picked up by the model, or bias from scorekeepers with respect to individual passers. 

\addtolength{\tabcolsep}{-3pt} 
\begin{table}
	\caption{NBA players with the top and bottom ten passer effects where ``Effect'' is the isolated effect of the passer on the recorded assist probability of an average assist (computed using Equation \ref{effect_computation}), and ``Rank'' is the rank of ``Effect'' in decreasing order (and ranges from 1 to 473)}
	\label{passer_effects}
	\begin{minipage}{.5\linewidth}
		\begin{tabular}{l l c c }
			\hline\noalign{\smallskip}
			\textbf{Rank} & \textbf{Passer Name} & \textbf{Position} & \textbf{Effect (\%)} \\
			\noalign{\smallskip}\hline\noalign{\smallskip}
			1 & Nick Collison & C & 14.42 \\ 
			2 & James Harden & SG & 13.29 \\
			3 & LeBron James & SF & 13.03 \\ 
			4 & Russell Westbrook & PG & 12.06 \\
			5 & Josh McRoberts & PF & 11.61 \\ 
			6 & G. Antetokounmpo & SF & 11.55 \\ 
			7 & Chris Paul & PG & 11.52 \\ 
			8 & Luke Babbitt & SF & 11.37 \\ 
			9 & Tony Allen & SF & 11.2 \\ 
			10 & Ricky Rubio & PG & 10.71 \\ 
			\hline\noalign{\smallskip}
			\end{tabular}
	\end{minipage}
	\vrule{}
	\begin{minipage}{.5\linewidth}
		\begin{tabular}{l l c c }
			\hline\noalign{\smallskip}
			\textbf{Rank} & \textbf{Passer Name} & \textbf{Position} & \textbf{Effect (\%)} \\
			\noalign{\smallskip}\hline\noalign{\smallskip}	
				473 & Andre Roberson & SG & -12.99 \\ 
				472 & Andre Drummond & C & -11.63 \\ 
				471 & Richard Jefferson & SF & -11.23 \\ 
				470 & Quincy Acy & PF & -10.75 \\ 
				469 & Tony Wroten & PG & -10.41 \\ 
				468 & Hassan Whiteside & C & -9.88 \\ 
				467 & Enes Kanter & C & -9.49 \\ 
				466 & Lavoy Allen & C & -9.30 \\ 
				465 & Andre Miller & PG & -9.21 \\ 
				464 & Hollis Thompson & SG & -9.07 \\ 
				\hline\noalign{\smallskip}
			\end{tabular}
		\end{minipage}
\end{table}
\addtolength{\tabcolsep}{3pt}

\subsection{Model Validation and Consistency}
\label{model_validation_and_consistency}

We wish to test the accuracy of our model in predicting whether a new potential assist will be recorded as an assist. To measure this accuracy, we use 10-fold cross validation to obtain mean log likelihood values and misclassification rates for the held out data. The misclassification rate is computed using the model as a classification tool with a cutoff of a probability of 0.5. We also compare the accuracy results of our model to those of three other models. That is, we compare the results for the full contextual model (Model \ref{contextual_model}), to the results for a model with no scorekeeper covariates, a model with no context covariates, and an intercept model. The model with no scorekeeper covariates is simply Model \ref{contextual_model} without the scorekeeper generosity and scorekeeper bias information. Comparing this model to the full model demonstrates whether the inclusion of the scorekeeper information improves the model. The model with no context covariates removes all information obtained from the optical tracking data, leaving Model \ref{simple_model}. Comparing this model to the full model demonstrates the performance differences between the existing methods and our new methods. Finally, the intercept model contains only an intercept term and no other covariates, and thus it treats every potential assist in an identical fashion (classifying each as a recorded assist). This model provides an estimate of the baseline accuracy to compare with the other models. 

The model validation results are presented in Table \ref{model_validation}. The results show that our new methods (Model \ref{contextual_model}) far outperform the previous best practice (Model \ref{simple_model}), leading to a misclassification rate of only 0.066. Additionally, by these metrics, the previous best practice had little to no improvement over simply using an intercept model. Finally, the results demonstrate a noticeable improvement with scorekeeper covariates included, even with all other covariates present. 

\begin{table}
	\caption{Out of sample mean log likelihood and misclassification rate results from a 10-fold cross validation performed on the set of all potential assists from the 2015-2016 season. The mean log likelihood values represent the average value for a single out of sample observation.}
	\label{model_validation}
	\begin{tabular}{l c c c}
		\hline\noalign{\smallskip}
		\textbf{Model} & \textbf{Mean Log Likelihood} & \textbf{Misclassification Rate} \\
		\noalign{\smallskip}\hline\noalign{\smallskip}
		Model \ref{contextual_model} (Full Contextual Model) & -0.176 & 0.066 \\
		No Scorekeeper Covariates & -0.182 & 0.070 \\
		Model \ref{simple_model} (No Context Covariates) & -0.638 & 0.344 \\
		Intercept Model & -0.644 & 0.344 \\
		\hline\noalign{\smallskip}
	\end{tabular}
\end{table}

We now wish to examine the stability of our model across NBA seasons. Optical tracking data is available in all stadiums for each of the seasons ending in 2014, 2015, and 2016 so we estimate models for each of these seasons and compare the resulting coefficient values. The results of this comparison are presented in Table \ref{model_comparison} in the form of correlation values across seasons for groups of coefficients. The results show that the coefficients which are common to all teams and scorekeepers (position and court zone effects) are all highly correlated across seasons. This result implies that these coefficients are detecting a meaningful signal, that remains consistent over time. Additionally, the scorekeeper bias and scorekeeper generosity effects display a moderate and a high level of correlation respectively. Again it appears as though these coefficients are detecting a meaningful and consistent signal. As previously discussed, the scorekeeper bias coefficients are estimated using less data than the generosity coefficients, likely leading to their slightly reduced correlation values. Finally both the opponent and team coefficients have little to no correlation across seasons. This result is expected since the players, coaches, and playing styles of a team are much more likely to change across seasons than the scorekeepers of a team. Thus, the consistency of the coefficients are highly team dependent. For example, across the three seasons, both the team (0.014, 0.020, and 0.068) and opponent (0.062, 0.046, -0.011) coefficients of the Boston Celtics have remained fairly consistent. This consistency may be explained by the system implemented by Brad Stevens, head coach of the Celtics across all three seasons. Conversely, in the 2014 off season, LeBron James announced his return to the Cleveland Cavaliers. In addition, the team also acquired Kevin Love through trade, and hired David Blatt to be its new head coach. This coaching change and addition of two all-star players drastically altered the playing style of the team, and this change was reflected in both the team (-0.280 and 0.326) and opponent (-0.012 and 0.217) coefficients across the seasons ending in 2014 and 2015.

\begin{table}
	\caption{Correlation of estimated coefficient values for a variety of effects across models estimated for the seasons ending in 2014, 2015, and 2016. Each estimated model follow the form of Model \ref{contextual_model}. The effects are sorted in increasing order by their 2014 and 2016 correlation values.}
	\label{model_comparison}
	\begin{tabular}{l r r r}
		\hline\noalign{\smallskip}
		\textbf{Effect} & \textbf{2014 and 2015} & \textbf{2015 and 2016} & \textbf{2014 and 2016} \\
		\noalign{\smallskip}\hline\noalign{\smallskip}
		Opponent & -0.065 & 0.003 & -0.164 \\
		Team & -0.284 & -0.001 & 0.204 \\
		Scorekeeper Bias & 0.486 & 0.631 & 0.460 \\
		Court Zone Interaction & 0.820 & 0.742 & 0.644 \\
		Passer Court Zone & 0.688 & 0.924 & 0.757 \\
		Scorekeeper Generosity & 0.887 & 0.856 & 0.776 \\
		Positions & 0.965 & 0.953 & 0.898 \\
		Shooter Court Zone & 0.920 & 0.990 & 0.903 \\
		\hline\noalign{\smallskip}
	\end{tabular}
\end{table}

\subsection{Adjusting Player Assist Totals}
\label{adjusting_player_assist_totals}
Since we have isolated the impact of each variable in the contextual model on the probability of a potential assist being a recorded assist and verified Model \ref{contextual_model} produces accurate prediction results, we can use the model to compute adjusted assist totals for each player. We compute adjusted assists using Equation \ref{effect_computation} to estimate the predicted probability of a pass being a recorded assist after removing the effects of all non-contextual variables (scorekeeper effects, position effects, etc.). As opposed to the adjustment methods of \cite{park_factors} and \cite{schuckers}, our method examines every potential assist individually. Our method also allows us to determine the expected number of recorded assists gained or lost by a player due to an individual factor, such as the passer effect for that player. The ten players with the greatest increase and the greatest decrease in total assists after this adjustment are presented in Table \ref{adjusted_results}. The results show that the ``Home Scorekeeper'' effect, which is the sum of all assists gained or lost due to the generosity and bias of the home scorekeeper for a player, tends to be the greatest potential deflater of the players' recorded totals. This observation is emphasized by the fact that seven of the ten players with the greatest increase is assists were members of the Utah Jazz, the team with the most negative scorekeeper bias coefficient. Examining the players whose assist totals decrease, the ``Home Scorekeeper'' effect is again often an important factor, as is the ``Passer'' effect. For point guards, the ``Position'' effect also contributed to the decreased totals. Conversely, the ``Away Scorekeeper'' effect, which is the sum of all assists gained or lost due to the generosity of the away scorekeepers for a player, tends to be relatively insubstantial across all players.

\addtolength{\tabcolsep}{-2.5pt} 
\begin{table}[ht]
	\caption{Comparisons of the total recorded and adjusted assists for the 10 players experiencing the greatest increases and the 10 players experiencing the greatest decreases. The adjusted totals are computed using Equation \ref{effect_computation} to compute the effects of all non-contextual variables and remove them from the recorded totals. The ``Assist Change'' column measures the difference between the recorded and adjusted totals. The ``Original Rank'' and ``Adjusted Rank'' columns provide the players' ranks (1-473) before and after the adjustment respectively. The four right-most columns display the estimated number of additional assists a player originally received due to the given effect (SK is short for scorekeeper) which were removed in the adjustment process. Note that not all factors in the adjustment are displayed, so the ``Assist Change'' column is not equal to the negative sum of the four displayed adjustment effects.}
	\label{adjusted_results}
	\begin{tabular}{ l  r  r  r  r  r  r  r  r }
		\hline\noalign{\smallskip}
		& \multicolumn{1}{c}{\textbf{Assist}} & \multicolumn{1}{c}{\textbf{Recorded}} & \multicolumn{1}{c}{\textbf{Original}} & \multicolumn{1}{c}{\textbf{Adjusted}} & 
		&  & \multicolumn{1}{c}{\textbf{Home}} & \multicolumn{1}{c}{\textbf{Away}} \\
		\textbf{Player} & \multicolumn{1}{c}{\textbf{Change}} & \multicolumn{1}{c}{\textbf{Assists}} & \multicolumn{1}{c}{\textbf{Rank}} & \multicolumn{1}{c}{\textbf{Rank}} & \multicolumn{1}{c}{\textbf{Position}} & \multicolumn{1}{c}{\textbf{Passer}} & \multicolumn{1}{c}{\textbf{SK}} & \multicolumn{1}{c}{\textbf{SK}} \\
		\noalign{\smallskip}\hline\noalign{\smallskip}
			Gordon Hayward & 28.38 & 287 & 46 & 36 & -0.26 & -2.80 & -24.90 & -0.17 \\ 
			George Hill & 18.64 & 257 & 55 & 48 & 3.98 & -4.53 & -20.60 & 0.25 \\ 
			Trevor Booker & 18.29 & 83 & 222 & 187 & -0.19 & -3.75 & -12.71 & -0.01 \\ 
			Rodney Hood & 16.83 & 208 & 71 & 67 & 0.41 & 0.25 & -16.74& -1.39 \\ 
			Joe Ingles & 16.74 & 93 & 203 & 171 & 0.29 & -4.35 & -11.28 & 0.58 \\ 
			Lavoy Allen & 12.93 & 76 & 234 & 214 & -1.60 & -3.88 & -6.03 & 0.71 \\ 
			Tyson Chandler & 12.88 & 64 & 262 & 233 & -2.09 & -4.11 & -7.06 & -0.63 \\ 
			Derrick Favors & 11.73 & 93 & 204 & 178 & -0.20 & 0.32 & -11.18 & -0.92 \\ 
			Trey Lyles & 10.16 & 60 & 273 & 242 & -0.11 & -1.33 & -8.22 & 0.35 \\ 
			P.J. Tucker & 9.83 & 176 & 96 & 81 & -0.19 & -2.34 & -9.03 & -1.62 \\ 
		\hline\noalign{\smallskip}

			Ish Smith & -19.60 & 495 & 11 & 12 & 4.99 & 1.13 & 11.78 & -0.66 \\ 
			James Harden & -22.63 & 602 & 6 & 6 & 0.82 & 16.37 & 6.36 & -2.65 \\ 
			Reggie Jackson & -24.34 & 483 & 14 & 15 & 5.79 & 6.76 & 10.66 & -0.22 \\ 
			Jrue Holiday & -27.22 & 386 & 20 & 22 & 4.56 & 2.28 & 14.11 & -1.74 \\ 
			LeBron James & -29.95 & 505 & 10 & 13 & -0.32 & 13.32 & 14.72 & -0.90 \\ 
			Tony Parker & -30.07 & 377 & 21 & 27 & 3.83 & 7.97 & 9.67 & -0.05 \\ 
			Draymond Green & -30.09 & 587 & 7 & 7 & -0.47 & 12.41 & 6.26 & -1.02 \\ 
			G. Antetokounmpo & -30.56 & 342 & 29 & 39 & -0.38 & 14.23 & 4.96 & 1.69 \\ 
			Ricky Rubio & -31.12 & 645 & 5 & 5 & 8.06 & 21.67 & -7.03 & -0.14 \\ 
			Chris Paul & -34.80 & 729 & 4 & 4 & 5.73 & 16.34 & 11.06 & -0.90 \\ 
	
		\hline\noalign{\smallskip}
		\end{tabular}
\end{table}
\addtolength{\tabcolsep}{2.5pt}

\section{Impact of Scorekeeper Inconsistency on Daily Fantasy Sports}
\label{daily_fantasy}
In daily fantasy contests individuals pay entrance fees, select a roster of players who generate fantasy points, and potentially receive a payout depending on the performance of their roster, all within the span of 24 hours. The popularity of such games is increasing, and so too is the amount of money at stake. In 2014, FanDuel Inc. and DraftKings Inc., currently the two largest daily fantasy operators in North America, together awarded over \$800 million in prizes across all sports and pledged to increase that number to over \$3 billion in 2015 \citep{okeeffe}. Since the NBA daily fantasy contest scoring systems for both companies rely exclusively on box score statistics (including assists and blocks) scorekeeper behaviour has significant influence on these scores. 

In addition to the overall bias or generosity of scorekeepers, the variability of their behaviour is also important to daily fantasy participants. Depending on their selection criteria and the contest they enter, a participant may either seek or avoid a player in a game whose scorekeeper has a high level of variability in the recording of statistics, since this variability affects the overall variability of a player's performance.

\begin{figure*}
	\centering
	\includegraphics[width=\textwidth]{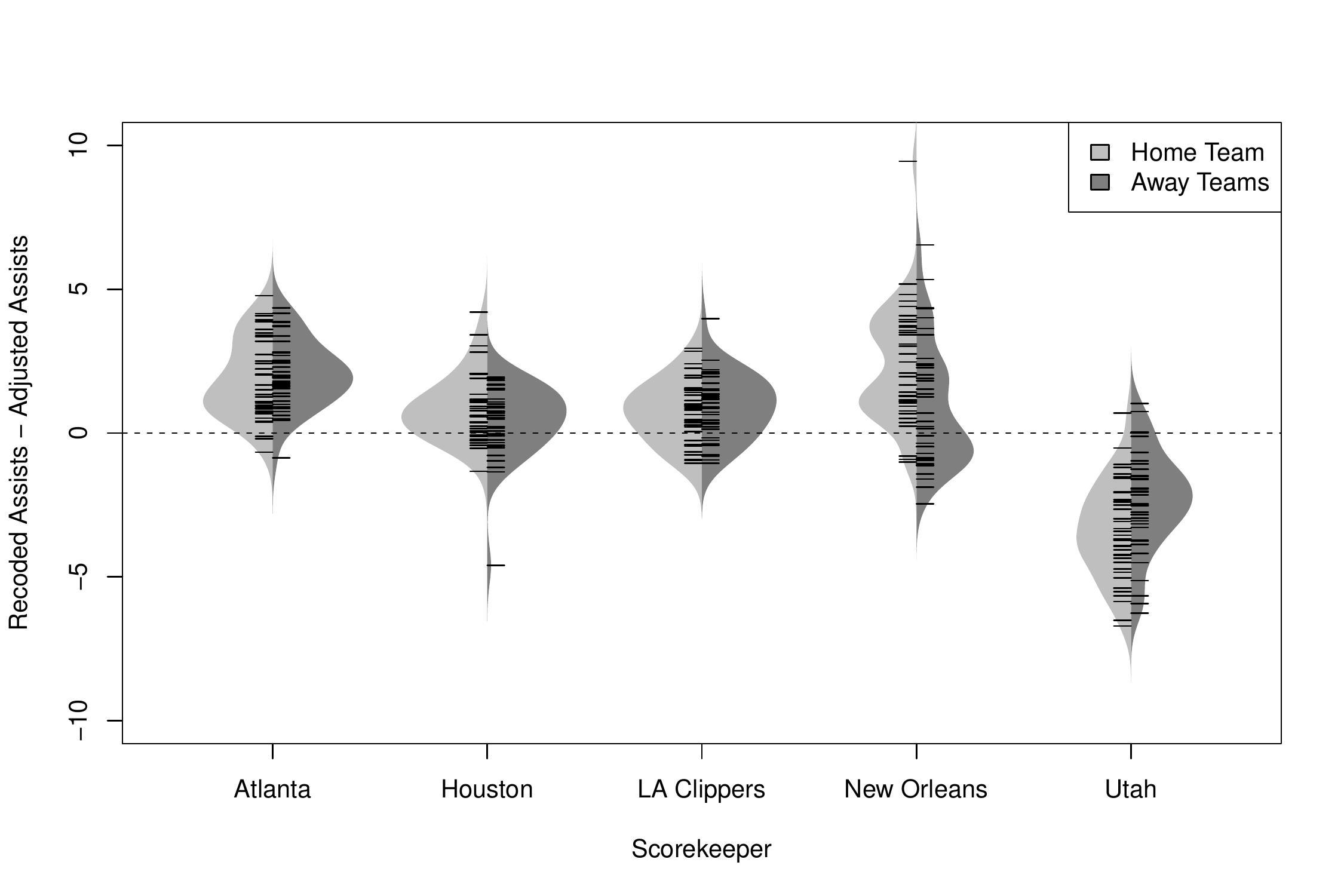}
	\caption{Scorekeeper bonus distributions of the home and away teams for 5 selected scorekeepers}
	\label{assist_beanplot}
\end{figure*}

Using the contextual model, we can compute adjusted assist totals for each team in all 1227 games in the data set by computing the sum of the expected probabilities of the potential assists after removing the estimated scorekeeper effects. These adjusted values represent the expected assist totals for each team in every game if an average scorekeeper had recorded the statistics. We can then compare the number of recorded assists to the number of adjusted assists and collect the difference values for both the home and away teams for each scorekeeper to obtain a home and away ``scorekeeper bonus" distribution for each of the 30 NBA scorekeepers. The home and away team distributions for 5 selected scorekeepers are presented in Figure \ref{assist_beanplot}. The means of the distributions range from -3.44 for the home team of the Utah Jazz scorekeeper to 2.32 for the home team of New Orleans Pelicans scorekeeper. Given that teams averaged 22.05 assists per game over the 1227 games in the data set, this difference of 5.76 assists per game is substantial. Both the home and away teams are more likely to get extra assists when playing in Atlanta, where almost all of both distributions are above zero, compared to Utah where with the majority of observations are below zero. Additionally, the Pelicans scorekeeper is the most unpredictable, with the greatest distribution variance values for both the home (4.03) and away (4.54) teams. However, not all scorekeepers are inaccurate and inconsistent. The scorekeeper for the Los Angeles Clippers is the most consistent in the league with a home distribution variance of 1.09 and an away distribution variance of 1.21 while the scorekeeper for the Houston was most accurate by the metric of mean absolute distance from zero with a home value of 1.06 and an away value of 0.997. 

\section{Discussion and Conclusion}
\label{conclusion}
In this paper we have presented evidence of inconsistencies in the awarding of box score statistics by the 30 team-hired scorekeepers. To quantify these inconsistencies, we used spatio-temporal data from optical tracking systems to develop a new model for predicting recorded assists. Our model was shown to have a greater predictive accuracy than previous methods, and also allowed us to develop an improved method of statistic adjustment. Though we only presented adjusted results for the contextual model from Section \ref{new_assists_model}, the results of the team-level model from Section \ref{team_adjusted} can also be used to adjust recorded assist and block totals.

In addition to demonstrating inconsistencies in the awarding of assists by the scorekeepers, both to all opposing teams and to their corresponding home teams, the results of the contextual model indicate scorekeepers may have biases in regard to both passer positions and the individual passers. Though this model attempts to include the coefficients we suspect have the greatest impact on the probability of a recorded assist, basketball is a complex system of positioning, events, and interactions, and it is impossible to include all potential factors in any model. As such, the difference in position and passer effects may be the result of characteristics that extend beyond the model. Further work must be completed in order to verify these results.

The same level of detail we used to examine assists could also be applied to the examination of other statistics. We have already presented evidence of scorekeeper inconsistencies in the recording of blocks, and the same may be true for other statistics such as rebounds or turnovers.

In addition to the inconsistencies among the 30 NBA scorekeepers, Section \ref{daily_fantasy} provides evidence of the inconsistency of individual scorekeepers among different games. These inconsistencies have real world consequences, including the rising potential of monetary consequences for daily fantasy participants due to the growing popularity of the contests. 

In light of the findings of this paper, the NBA and its players may be well served to adopt a more proactive stance towards monitoring the attribution of subjective box score statistics. While our approach provides an adjustment for players’ assist totals, significant work remains to understand the impact scorekeeper inconsistencies have on aggregate metrics (such as PER, and WS) and on the salaries, perception, and careers of players.

\appendix
\section{Availability of Data}
Section \ref{assists_and_blocks} and Section \ref{team_adjusted} use ESPN box score data from the 2015-2016 NBA season. This data is publicly available and can be found at \url{http://www.espn.com/nba/scoreboard}. The SportVu optical player tracking data from STATS LLC used in Section \ref{new_assists_model} for the 2013-2014, 2014-2015, and 2015-2016 NBA seasons remains proprietary. However, to address concerns of reproducibility, our lab has released a full game of tracking data, available at \url{https://github.com/dcervone/EPVDemo/blob/master/data/2013_11_01_MIA_BKN.csv}.

\bibliographystyle{plainnat}

\end{document}